\documentclass[sigconf]{acmart}

\usepackage{colortbl}
\usepackage{framed}
\usepackage{multirow}
\usepackage{rotating}
\usepackage{soul}
\usepackage{xcolor}

\newcolumntype{R}[2]{%
    >{\adjustbox{angle=#1,lap=\width-(#2)}\bgroup}%
    l%
    <{\egroup}%
}

\copyrightyear{2026}
\acmYear{2026}
\setcopyright{cc}
\setcctype{by}
\acmConference[ICSE-Companion '26]{2026 IEEE/ACM 48th International Conference on Software Engineering}{April 12--18, 2026}{Rio de Janeiro, Brazil}
\acmBooktitle{2026 IEEE/ACM 48th International Conference on Software Engineering (ICSE-Companion '26), April 12--18, 2026, Rio de Janeiro, Brazil}
\acmPrice{}
\acmDOI{10.1145/3774748.3795685}
\acmISBN{979-8-4007-2296-7/2026/04}

\begin{document}

\title{Towards Understanding Views on Combining Videos and Gamification in Software Engineering Training}

\author{Pasan Peiris}
\affiliation{%
  \institution{University of Canterbury}
  \country{New Zealand}
  }
\email{pitigalagepasan.peiris@pg.canterbury.ac.nz}

\author{Matthias Galster}
\affiliation{%
  \institution{University of Bamberg \\ University of Canterbury}
  \country{Germany, New Zealand}
  }
\email{mgalster@ieee.org}

\author{Antonija Mitrovic}
\affiliation{%
  \institution{University of Canterbury}
  \country{New Zealand}
  }
\email{tanja.mitrovic@canterbury.ac.nz}

\author{Sanna Malinen}
\affiliation{%
  \institution{University of Canterbury}
 \country{New Zealand}
  }
\email{sanna.malinen@canterbury.ac.nz}

\author{Raul Vincent Lumapas}
\affiliation{%
  \institution{University of Canterbury}
  \country{New Zealand}
  }
\email{raulvincent.lumapas@pg.canterbury.ac.nz}

\author{Jay Holland}
\affiliation{%
  \institution{University of Canterbury}
  \country{New Zealand}
  }
\email{jay.holland@canterbury.ac.nz}

\renewcommand{\shortauthors}{Peiris et al.}


\begin{abstract}
Watching training videos passively leads to superficial learning. Adding gamification can increase engagement. We study how software engineering students and industry practitioners view gamifying video-based training. We conducted a survey with students and professionals. Students and professionals share similar perceptions toward video-based training in general and support combining gamification and video-based training. Our findings can inform the design of gamified training solutions for software engineers.
\end{abstract}

%


\maketitle

\section{Introduction and Background}

Software engineering training involves educating future professionals and up-skilling current professionals.
There are, however, several differences between students and professionals, e.g., training objectives and learner characteristics. 

Video-based training has become popular, allowing learners to learn at their own pace~\cite{sablic2021video}, e.g., training delivered via Coursera, LinkedIn Learning or YouTube tutorials. Previous works showed effectiveness and appreciation of video-based training with software engineering students~\cite{shynkarenko2024video} and professionals~\cite{okano2018enhancing}. However, passive video-based training leads to low retention of knowledge~\cite{chatti2016video}. Therefore, in \emph{active video watching} viewers actively and consciously engage in the educational material via interactive elements (e.g., commenting on videos and reviewing others' comments)~\cite{dimitrova2022choice}. Integrating gamification can further improve learner motivation and training effectiveness~\cite{de2017gamification}. Our goal is to understand perceptions of software engineering students and professionals about gamification to support video-based training.

\section{Research Method}
\label{Research Method}
We conducted a survey and recruited 85 students from two universities. We recruited 100 professionals via Prolific. We screened professionals based on several criteria and screening questions. We used AVW-Space~\cite{lau2016usability} as the video-based training platform (not gamified) and customized it for presentation skills, one representative communication skill. The platform includes three generic activities: watching videos, commenting on videos, and reviewing comments made by other learners. Participants used the platform for 40 minutes and then watched an introductory video on gamification. Then, participants answered a questionnaire. We analyzed closed questions using descriptive statistics. We analyzed response frequencies and cross-tabulated answers across questions. We analyzed open-ended questions using content analysis~\cite{seidel1998qualitative}.
\section{Results}
\label{sec:results}

\textbf{Hardest Activity:}
Most students (57\%) and professionals (60\%) found commenting on videos to be the hardest activity. This aligns with the prior studies which found that commenting on videos requires deep intellectual engagement. A Chi-square test comparing students and professionals revealed no significant difference.

\noindent\textbf{Least Motivating Activity:}
Both groups found reviewing comments the least motivating activity (59\% of both students and professionals). Previous studies also indicate that learners found reviewing comments to be frustrating.
A Chi-square test comparing students and professionals revealed no significant difference.

\noindent\textbf{Most Useful Activity:}
More than two-thirds of students (70\%) and  professionals (79\%) found watching videos most useful for learning. A Chi-square test comparing revealed no difference.

\noindent\textbf{Motivating Gamification:}
We asked participants to name gamified platforms before providing reasons for why gamification was motivating. Forty-eight students did not name any motivating platform. However, 38 students gave examples (e.g., Duolingo, Kahoot, Education Perfect) and reasons. We identified three key reasons: (1) competitive nature, (2) accomplishments through rewards, (3) visualization of progress. Among professionals, 39 indicated no prior motivating experiences with gamified platforms, while 63 provided examples (e.g., Duolingo, Kahoot, Codingame) and reasons.
The primary reasons were: (1) fun and enjoyable nature of the platform, (2) sense of accomplishment by rewards, and (3) visualization of progress. Interestingly, both students and professionals found visualization of progress and sense of accomplishment as motivating. On the other hand, professionals enjoy the fun nature of gamification, while students appreciate competition.

\noindent\textbf{Annoying Gamification:}
We asked participants to name platforms when explaining why gamification can be annoying. Some platforms were cited as both motivating and annoying. Fifty-seven students reported no annoying experiences, while 29 gave examples (e.g., Duolingo, Education Perfect) and reasons. The top 3 reasons for why students found these platforms annoying were: (1) poor gamification design, (2) technical problems, and (3) excessive advertisements. Among professionals, 68 participants reported no annoying experience, while thirty-five identified examples (e.g., Duolingo, Kahoot) and reasons: (1) restrictive or coercive design, (2) excessive notifications, (3) unclear learning paths and unrealistic goals. Duolingo, Kahoot, and Education Perfect also ranked highly among motivating platforms. Reasons for why gamified platforms are considered annoying differ between professionals and students. Individual preferences, expectations, and backgrounds (e.g., age, learning goals and setting) and the skills taught on these platforms may lead to distinct motivations and frustrations.
\section{Discussion}

Both groups considered the  three generic activities in video-based training (watching, commenting, reviewing comments) similarly regarding difficulty, usefulness, and motivation. Hence, we should enhance \emph{all} activities with gamification rather than only \emph{one} of these activities. For instance, gamification could support the least motivating activity (reviewing comments) while further improving engagement in the most useful one (watching videos). However, we also need to consider the impact of gamification on learning. Students are used to structured, competitive, and exam-driven learning environments, and may therefore associate gamification with motivation and improved learning. Also, they may have encountered gamification in their academic journey already, making them more receptive. In contrast, professionals may engage in self-directed, goal-oriented learning tailored to their specific job roles. While gamification can increase motivation and engagement, professionals may be more skeptical about whether these factors directly translate into better skill development. Professionals may even perceive gamification more as entertainment rather than serious training enhancement. Furthermore, professional training prioritizes practical application, but it is difficult to provide hands-on experience of soft skills via video-based training, which again might impact views on gamifying video-based training.

Both students and professionals identified flawed gamification design as a source of frustration. This underscores the need to understand theoretical and conceptual foundations of gamification to carefully design solutions. To enhance training using gamification, designers must justify their choice of mechanics based on established theories rather than arbitrary decisions. Our study contributes to this understanding by identifying the shared motivators and frustrations of both students and software professionals, offering a foundation for more effective gamification design in training environments for software engineering domain. For example, to avoid user frustration, designers should avoid excessive notifications and ensure that goals to achieve rewards are clear and realistic.

Features such as progress visibility and reward systems emerged as motivating for both students and professionals. This highlights the importance of personalization when designing gamified solutions. Interestingly, students were primarily motivated by the competitive nature of gamification platforms, likely due to the competitive dynamics of classroom settings. In contrast, software professionals were more driven by the fun aspect of gamification.

We acknowledge the potential issue of only asking about perceptions, which might not well reflect the actual behaviour in the presence of gamification. Also, there is a possibility that different individuals might react differently to gamification choices (e.g., some might like competition, others not), which complicates the design of gamification.
\section{Conclusion}

Students and  professionals share similar views on video-based training. Both groups generally support integrating gamification into video-based training, though professionals are more skeptical about its impact on learning. We used a specific video-based training platform and a specific skill to guide responses, but watching, commenting, and reviewing are generally applicable to all platforms. Cultural differences between countries may have influenced participants' perceptions, so a potential limitation is the uneven distribution of participants across different regions. Our insights can help researchers and gamification designers develop effective design principles and mechanisms for integrating gamification into (video-based) software engineering training.

\bibliographystyle{ACM-Reference-Format}
\bibliography{references}

\end{document}